# Direct observations of cross-scale energy transfer in space plasmas

Jing-Huan Li[1], Xu-Zhi Zhou[1*], Zhi-Yang Liu[1], Shan Wang[1], Yoshiharu Omura[2],
Li Li[1], Chao Yue[1], Qiu-Gang Zong[1], Guan Le[3], Christopher T. Russell[4], James L. Burch[5]

1. School of Earth and Space Sciences, Peking University, Beijing, China
2. Research Institute for Sustainable Humanosphere, Kyoto University, Kyoto, Japan.
3. NASA Goddard Space Flight Center, Greenbelt, MD, USA.
4. Institute of Geophysics and Planetary Physics, University of California, Los Angeles, CA, USA.
5. Southwest Research Institute, San Antonio, TX, USA.

\* Corresponding author: Xu-Zhi Zhou (xzzhou@pku.edu.cn)

**Abstract**

The collisionless plasmas in space and astrophysical environments are intrinsically multiscale in nature, behaving as conducting fluids at macroscales and kinetically at microscales comparable to ion- and/or electron-gyroradii. A fundamental question in understanding the plasma dynamics is how energy is transported and dissipated across different scales. Here, we present spacecraft measurements in the solar wind upstream of the terrestrial bow shock, in which the macroscale ultra-low-frequency waves and microscale whistler waves simultaneously resonate with the ions. The ion acceleration from ultra-low-frequency waves leads to velocity distributions unstable to the growth of whistler waves, which in turn resonate with the electrons to complete cross-scale energy transfer. These observations, consistent with numerical simulations in the occurrence of phase-bunched ion and electron distributions, also highlight the importance of anomalous resonance, a nonlinear modification of the classical cyclotron resonance, in the cross-scale wave coupling and energy transfer processes.

**Main**

In the plasma universe, the energy and momentum exchange usually cannot be supported by collisions as in an ordinary gas, but rather rely on the variety of electromagnetic and electrostatic waves. The coexistence of many different wave modes, together with the diverse species and energies of charged particles, constitute a highly-coupled system where cross-scale processes occur through turbulent cascades and/or resonant wave-particle interactions [1]. A basic pattern of the latter process is that the particle distributions can be modulated by low-frequency waves and become unstable to excite plasma waves at higher frequencies [2-4], which could be eventually absorbed through interactions with other particles.

The above processes, although thought to be universal, are usually difficult to observe. A unique natural laboratory enabling in situ investigation of such processes is the near-Earth space, especially in recent years when high-resolution particle and field measurements are available via spaceborne instruments. A region of particular interest is Earth's foreshock, in which charged particles backstreaming from the terrestrial bow shock interact with the background solar wind to accommodate a zoo of plasma waves [5,6]. One of the commonly observed wave modes is the fast-mode magnetosonic waves, typically in the ultra-low-frequency (ULF) range after being Doppler-shifted into the spacecraft rest frame [5]. These ULF waves are excited by the backstreaming ions via electromagnetic ion/ion resonant or nonresonant instabilities [7]. The nonlinear steepening of the ULF waves leads to the formation of sharp magnetic bumps known as "shocklets" [8,9], which are typically accompanied by magnetosonic-whistler wave packets radiating sunward in the plasma rest frame [5,10,11]. Similar signatures have been also reported near interplanetary shocks [12] and other planetary [13-15] or cometary environments [16]. These waves are believed to play a critical role in the energy exchange with solar wind particles [12,17-20], although the detailed processes responsible for the cross-scale energy transfer are still unclear due to the lack of direct observations.

In this paper, we present observations from NASA's Magnetospheric Multiscale (MMS) spacecraft of large-amplitude electromagnetic waves in the Earth's foreshock, providing direct evidence for cross-scale energy transfer. The observations show that the excitation of ULF waves by the reflected ions through the linear ion/ion resonant instability. As the ULF waves grow, the ions undergo simultaneous resonant interactions with large-amplitude ULF and whistler waves. The former resonance produces a new population on top of the background solar wind, leading to the growth of the whistler waves and the energy transfer from the fluid scale to the ion-gyration scale. The energy is then transferred to the electron-gyration scale through whistler wave-electron resonance. These observations support the scenario of cross-scale energy transfer via resonant interactions between particles and various plasma waves. The results also indicate that anomalous resonance, a nonlinear effect when the wave amplitude is large enough to modify the classical resonance condition[21], plays an important role in the energy transfer in regions of weak magnetic field.

**Results**
**Overview**
The event occurred on January 7, 2021 when the four-spacecraft MMS constellation traveled in the foreshock solar wind, at [23.7, 15.4, 0.3] Earth Radius (Re) in Geocentric Solar Eclipse (GSE) coordinates. The MMS data used here include the electric field measurements from Electric field Double Probes (EDP), magnetic field measurements from Flux Gate Magnetometer (FGM), and the ion/electron velocity distributions from Fast Plasma Investigation (FPI). These observational data with unprecedented high resolution offer a rare opportunity to study the wave-particle interactions quantitatively [22,23].

Figure 1 presents an 18-min overview of the MMS1 observations, in which the magnetic field components in GSE coordinates (Figure 1a) show strong oscillations (most significantly in the yz plane) with a period of ~45 s. The ULF oscillations of the field strength (black line in Figure 1b) are mostly in phase with the oscillations of the plasma density (blue line in Figure 1b), which is an important feature of the fast-mode magnetosonic waves commonly observed in the foreshock solar wind [24,25]. The background field, computed via a long-term average of the magnetic field, is almost always in the anti-sunward direction (or equivalently, the solar wind flow direction), with an amplitude of ~3nT comparable to the wave-carried magnetic field.

Therefore, this is a case with nearly radial IMF and the entire dayside upstream region is in the ion foreshock. The GSE coordinates could be approximately treated as the background field-aligned coordinate system, in which the parallel direction and the perpendicular plane are simply the -x direction and the yz plane, respectively. To facilitate the analysis on particle distributions, we accordingly define the pitch angle of any given particle as the angle between its velocity and the -x direction, and its phase angle as the angle between its perpendicular velocity $v_\perp$ (in the yz plane) and the -z direction.

Figure 1c shows the wavelet power spectrum of the transverse magnetic perturbations (the $B_y$ and $B_z$ components), in which the persistence of the large-amplitude ULF waves is manifested by an enhanced power spectral density near the frequency of 0.02 Hz. The wavelet spectrum also shows intermittent enhancements of the wave power at frequencies above 0.1 Hz. These high-frequency wave packets, with two examples shown in Figures 1g and 1j, are accompanied by sharp bumps of the magnetic field. These are characteristic signatures of nonlinearly steepened ULF waves associated with whistler precursors at higher frequencies [10]. The detailed wave properties will be given below.

**Wave properties**

We first focus on the first shadowed interval in Figure 1 (between 0821:30 and 0824:30 UT). The ULF wave field in the yz plane denotes a left-hand polarization, since the quasi-sinusoidal $B_y$ variations always lead in phase than $B_z$ by ~π/2. After this time interval, the ULF waves gradually become linearly polarized in association with the appearance of sharp magnetic bumps (see Figure 1a, after 0825:00) [8].

We next estimate the phase speed of the ULF waves. Since the Poynting flux is predominantly in the -x direction (along the bulk flow direction), the parallel phase speed $v_{ULF}$ can be determined based on Faraday's law [21,26] to be ~450 km/s. This speed is slightly lower than the plasma bulk velocity $v_p$ (465±9 km/s along the -x direction), which suggests that the left-hand polarized ULF waves in the spacecraft frame could be right-hand polarized in the plasma rest frame [5], although the similar values between $v_{ULF}$ and $v_p$ indicate a large uncertainty in the determination of the Doppler-shifted wave properties.

The properties of the high-frequency wave packets are also analyzed. The two typical time intervals (0825:00~0825:20 and 0831:30~0831:50 UT), with zoomed-in views given in Figures 1g-1i and 1j-1l, are both characterized by steepened bumps in magnetic field strength associated with large-amplitude wave packets (see Figures 1g and 1j), although they also show distinctly different properties. During the first interval, the field strength and the wave period both increase with time, whereas the opposite trends occur during the second interval. Moreover, the high-frequency waves (filtered between 0.2Hz and 3Hz, see Figures 1i and 1l) are right-hand and left-hand polarized during these time intervals, respectively.

The different signatures can be also understood by the different phase velocities of the high-frequency waves during these intervals. Based on the inter-spacecraft phase delay (see ref.[18] for detailed description of the approach), we determine the wave phase speeds of 640±30 km/s and 407±7 km/s along the background flow direction during the first and the second intervals, respectively. Their higher and lower values than the bulk flow speed (465±9 km/s ) indicate the earthward and sunward propagation of the waves in the plasma rest frame, respectively, which leads to different observation sequences between the leading and the trailing edges (see Figures 1g and 1j). The lower phase speed during the second time interval also indicates a polarization reversal due to the Doppler effect. In other words, the waves in both intervals are right-hand polarized in the plasma rest frame, which supports the whistler-mode nature of these waves. The frequency dispersion originates from the larger phase speed of the higher-frequency waves than lower-frequency waves [27], so that the leading edge always corresponds to higher frequencies. The different wave propagation directions during these two intervals, on the other hand, indicate that the whistler waves are excited near the spacecraft location[28,29].

**ULF wave interaction with the ions**

Figures 1d-1f are overviews of the ion distributions and their correlations with the ULF waves. Figure 1d shows the ion pitch-angle spectra within the energy range between 1.5 and 4.5 keV, in which the high phase-space densities (PSDs) in the parallel direction (pitch angle ~0°) correspond to the intense solar wind flow. During the first two minutes, there is another population at the pitch-angle of ~135°, which are reflected beams from the terrestrial bow shock. The reflected population can be also seen in Figure 2a, the averaged ion distributions in the $v_\parallel - v_\perp$ plane from 0816:00 to 0817:00. This population is mainly composed of protons, since the $He^+$, $He^{++}$, and $O^+$ ions are mostly concentrated in the -x direction (not shown). The reflected protons have long been associated with the ULF wave excitation in the foreshock region through the ion/ion resonant instability[7], which can be examined by considering the classical cyclotron resonance condition below.

The parallel speed of a cyclotron resonant particle, often called the resonance velocity, equals

$$V_r = \frac{(\omega \pm \Omega)}{k_\parallel}, \qquad (1)$$

where $\omega$ is the wave angular frequency, $\Omega$ is the particle gyrofrequency, and $k_\parallel$ is the parallel wave number. Here, the + and – signs correspond to the cases with the wave polarization having the opposite and the same handedness as the particle gyration, respectively[30]. In this event, the ULF waves are left-hand polarized in the spacecraft rest frame, and therefore, the minus sign applies for resonant ions. Moreover, the observed ULF wave frequency $\omega_{ULF}$ (~0.14 rad/s) is smaller than the proton gyrofrequency $\Omega_p$ (~0.29 rad/s), which indicates a negative $V_r$ of $-482$ km/s (see the purple dashed line in Figure 2a). For a resonant proton at 3keV, the pitch angle would be 130°, which belongs to the reflected beam population.

Based on the observed particle distributions (Figure 2a), we utilize a dispersion relation solver[20,31] (with the detailed parameters given in Methods section) to confirm the linear excitation of fast-mode magnetosonic waves propagating sunward in the plasma rest frame. Figure s1 shows the resultant wave dispersion relation and growth rate, in which the sunward wave propagation is manifested by the negative $k_{pl}$, the parallel wave number in the plasma rest frame (opposite to the case in the spacecraft frame). As we have discussed above, the reversed sign of $k_{pl}$ (while the wave frequency in the plasma rest frame $\omega_{pl}$ remains positive) is consistent with a right-hand wave polarization, indicating the adoption of the plus sign in equation (1). Figure 2c shows a schematic illustration of the wave excitation process in the plasma rest frame, in which the resonance condition (solid blue line, note that its expression is derived from equation (1) with the plus sign) intersects the wave dispersion relation (red line) at small $\omega_{pl}$ and $k_{pl}$ values indicating the long wavelength of the excited ULF waves[24,30].

As the ULF waves grow, the reflected beams gradually disappear, and there emerges another population in the pitch-angle range between 40° and 90° (Figure 1d). Note that the new population is well separated from the solar wind flow with pitch angles close to 0°. Figure 1d also shows the periodic variations of the ion PSDs at approximately the ULF wave frequency within the 1.5 ~ 4.5keV energy range. The wave-particle correlation can be better visualized in Figure 1e, which displays the energy spectrum of the ions moving in the -z direction superimposed by the magnetic field $B_z$ component (the black lines). Obviously, the periodic ion flux enhancements have a one-to-one correlation with the $B_z$ peaks. We further show in Figure 1f the wavelet coherence between the 3keV-ion flux and the $B_z$ field [32], with the arrows representing their phase differences. The high coherence at the wave period of ~45s, together with the horizontal arrows indicating the wave-particle in-phase relationship, demonstrate the phase-bunched ion distributions characterizing the occurrence of wave-particle resonance. Based on previous studies[1,21,26], the phase-bunched feature can be also manifested as the periodic, inclined stripes in the gyro-phase spectra if the waves are circularly polarized. Therefore, we select the first shadowed interval in Figure 1 (during which the waves are left-handed and nearly circularly polarized) to further analyze the ion distributions.

Figures 3b-3d show the gyro-phase spectra for 3keV ions (averaged over MMS1, MMS2 and MMS3 spacecraft to improve the statistical significance) at the pitch angles of 40°, 90° and 130°, respectively. The occurrence of periodic, inclined stripes is shown in Figures 3b and 3c, with enhanced ion PSDs in antiphase with the ULF wave field $\mathbf{B}_{1,\mathrm{ULF}}$ (see the white lines for the phase of $-\mathbf{B}_{1,\mathrm{ULF}}$). This feature is a manifestation of wave-particle resonance, which can only be expected within finite energy and pitch-angle ranges where the ions are locked in phase with the waves during their gyromotion (so that sustained acceleration may occur)[26]. As we have discussed before, the classical picture of cyclotron resonance indicates the resonance velocity of $V_r$ =-482 km/s, which corresponds to the pitch angle of 130° for 3keV-protons. This expectation contradicts with the phase-bunched signatures at the pitch angles of 40°~90° (Figures 3b-3c) but not 130° (Figure 3d). Even if we consider a large uncertainty in wave frequency (0.12~0.18 rad/s), the corresponding $V_r$ values would range from -600 to -300 km/s, still deviating significantly from the observations. In other words, the classical theory of cyclotron resonance may not be appropriate during this interval with large wave amplitude.

To understand the observations, we invoke the scenario of anomalous resonance[21] for waves with large amplitude comparable to the background magnetic field $B_0$ ($|B_{1,\mathrm{ULF}}|/|B_0|\sim 2/3$ in this event). In this scenario, the particle angular velocity can be significantly modified by the wave-associated forces, and consequently, the resonance condition is nonlinearly revised[33]. This scenario also indicates the occurrence of anomalous resonance at two separated phase-space locations [34,35], which are in phase and in antiphase with the wave magnetic field, respectively. The full resonance velocity is expressed by

$$V_r' = \frac{(\omega - \Omega)v_\perp \mp \Omega_1 v_w}{k_\parallel v_\perp \mp \Omega_1} = V_r \pm \frac{(V_r - v_w)}{\frac{k_\parallel v_\perp}{\Omega_1} \mp 1}, \qquad (2)$$

where the upper and lower symbols correspond to the in-phase and in-antiphase resonant islands, respectively. Here, $v_w$ represents the wave phase speed, and $\Omega_1 = B_1 q/m$ is the nominal gyrofrequency associated with the wave field $B_1$. Note that Equation (2) is degenerated into the classical resonance condition (1) when the following criterion is satisfied [21],

$$\sqrt{\frac{B_1}{B_0}} \ll \sqrt{\frac{v_\perp^3}{|v_w - V_r|^3}}. \tag{3}$$

Given a typical proton (3keV at the pitch angle of 45°) in this event, the left and right sides are 0.82 and 0.44, respectively. In other words, condition (3) is not satisfied due to the large $B_1/B_0$ ratio, which explains the discrepancy between the observations and the expectation from the classical resonance theory. Therefore, the anomalous resonance must be taken into account. Since the ions in Figures 3b-3d are phase bunched in antiphase with the wave magnetic field, we choose the lower sign in Equation (2) to calculate the full resonance velocity $V_r' = -59$ km s$^{-1}$ for the 3keV ions, which corresponds to the resonant pitch angle of 95° (rather than 130° based on Equation (1), also see Figure s2a for the resonant pitch angle at different energies). This is consistent with the observations that the most prominent phase-bunched signatures appear at the pitch angle of 90° for the 3keV ions. At pitch angles below 90° (Figure 3b), the resonant trapping is also made possible by the upward shift of the resonant island center (from -482 to -59 km s$^{-1}$), especially after the finite width of the resonant island is considered. This scenario is further verified via test-particle simulations.

The simulation follows a similar procedure as in ref.[21], which is based on Liouville's theorem to ensure the PSD conservation of ions along their trajectories [36-38]. The initial ion distributions utilized are the averaged distributions measured during the quiet period from 0816:00 to 0817:00. We next launch a Gaussian-profiled wave packet propagating along the homogeneous background field, and trace from an immobile virtual spacecraft a series of test ions backward in time to obtain their initial velocities. The corresponding PSDs are thus determined via Liouville's theorem. The detailed simulation setup and parameters are given in the Method section. Figures 3e-3h show the virtual spacecraft observations of the ULF waves and the ion gyro-phase spectra at the pitch angles of 40°, 90° and 130°, respectively. They all display periodic, inclined stripes with enhanced PSDs in antiphase with the wave field $\mathbf{B}_{1,\mathrm{ULF}}$, with larger amplitudes of the PSD variations at 40° and 90° than at 130°. These features are similar to the actual observations in Figures 3a-3d. Detailed analysis on typical ion trajectories indicates their trapping motion around the resonant island centered at $\zeta =180°$ (where $\zeta$ is the phase difference between $v_\perp$ and $B_1$, see the blue line in Figure s3a), during which their energy can be significantly increased (see Figure s3b). It is the acceleration process that causes the periodic PSD enhancement in antiphase with the wave field, which is also consistent with the observations of the enhanced PSDs in the pitch angle range between 40° and 90° (see Figure 1d). The consistency between the observations and the simulation results indicates the efficient acceleration of the ions via anomalous resonance as the wave amplitude becomes larger (especially after 0818:00).

**Whistler wave interaction with the ions**

We next study the interactions between the ions and the high-frequency, whistler wave packets. Taking the second shadowed interval (0831:30~0831:50 UT) in Figure 1 as an example, the whistler wave properties determined above enable us to examine criterion (3) for the applicability of the classical cyclotron resonance. Here, the classical resonance velocity $V_r$ ~ 372 km/s is close to the whistler wave phase velocity $v_{wh}$ ~ 400 km/s along the parallel direction in the spacecraft frame, which indicates the satisfaction of condition (3) and the applicability of the classical cyclotron resonance. Interestingly, the ions in resonance with the whistler waves overlap with those experiencing anomalous resonance with the ULF waves. Figure 2b shows the averaged ion distributions during this time interval (within the $v_\parallel - v_\perp$ plane), in which the area surrounded by the solid black box (with energy 1.5~4.5 keV and pitch angle 40°~90°) represents the ions in anomalous resonance with the ULF waves. The ion PSDs within this area are higher than those in adjacent phase-space regions to form an unstable distribution, which probably results from the sustained ion acceleration from the ULF waves. The purple line, on the other hand, delineates the resonance velocity with the whistler waves (see Figure s2b for the condition of cyclotron resonance). In other words, these ions could resonate simultaneously with the ULF and the whistler waves, and it is natural to speculate that the unstable ion distributions driven by the ULF waves could lead to the growth of the whistler waves.

We next calculate the energy transfer rate $\mathbf{J}_i \cdot \mathbf{E}_1$[4,26,39] between the ions and the whistler waves. Here, the ion current density $\mathbf{J}_i$ in the plasma rest frame represents the contribution from the resonant ions within the surrounded box in Figure 2b, and a Lorentz transformation is carried out to ensure that the electric field $\mathbf{E}_1$ is also in the plasma rest frame. Figure 3n shows the 4s-running-averaged energy transfer rate, which is predominantly negative to indicate the energy transfer from resonant ions to whistler waves. The transfer rate maximizes at ~ 0.015 pW m$^{-3}$, suggesting that it may take 2 gyro periods for the whistler waves to reach the amplitude of $B_1 = 1.5$ nT, close to previous simulation results [8]. The energy transfer, according to the nonlinear theory of cyclotron resonance[40], could originate from the nonzero inhomogeneous factor $S$ contributed either by the positive sweeping of the whistler waves or by the curvature of the background magnetic field. The whistler wave excitation can also be interpreted in Figure 2c as the intersection of the resonance condition (dashed blue line, with smaller beam velocity than the reflected beam in the plasma rest frame) and the dispersion relation at a shorter wavelength. Note that in this event, the wavelengths for the ULF and the whistler waves are around 2*10$^4$ and 10$^3$ km, respectively. Therefore, the wave-particle interactions provide an efficient way for transferring energy from macro-scales (165$r_i$, $k_{ULF}r_i = 0.04$) to micro-scales (8$r_i$, $k_{wh}r_i = 0.86$), where $r_i = 121$ km represents the ion inertial length. Here, $k_{ULF}$ and $k_{wh}$ are the parallel wave numbers in the spacecraft rest frame for the ULF wave and whistler wave, respectively.

**Whistler wave interaction with electrons**

We further analyze the interaction between the electrons and the whistler waves. Figures 3i-3m show the observed wave magnetic field and electron gyro-phase spectra (averaged over MMS1, MMS2 and MMS3 observations, since the MMS4 data is unavailable), in which the PSDs of electrons with the pitch-angles of 40° (Figures 3j-3k) and 140° (Figures 3l-3m) display periodic, inclined stripes in phase and in antiphase with the whistler wave magnetic field $\mathbf{B}_{1,wh}$ (see the white lines for the phase of $-\mathbf{B}_{1,wh}$), respectively. These features indicate the coexistence of two resonant islands in phase and in antiphase with $\mathbf{B}_{1,wh}$, a manifestation of anomalous resonance between electrons and whistler waves[21]. The corresponding anomalous resonance conditions (with

$\zeta = 0°$ and 180°, respectively) are shown in Figure s2c, which agree with the observations in Figure 3j-3m. In fact, one may determine the classical cyclotron resonance velocity based on equation (1) and the wave parameters given in the Methods section, to be $V_r = -4.93 * 10^4$ km/s, which corresponds to the parallel energy of ~7 keV. This energy is much larger than the ones in Figures 3j-3m (17-67 eV), which is consistent with the fact that criterion (3) of classical cyclotron resonance is unsatisfied.

To further validate the hypothesis of anomalous resonance occurrence in this event, test-particle simulations following similar procedures are carried out for the electrons, with the wave parameters and particle distribution given in the Methods section. The simulation results in Figures 3p-3s reproduce the signatures of electrons bunched in-phase or in-antiphase with the wave magnetic field. The typical electron trajectories (Figures s3c-s3d) also suggest the trapping motion of the resonant electrons, during which they can be accelerated by the whistler waves. The energy conversion occurs within the electron cyclotron scale, the order of ~10 km ($4r_e$, $k_{wh}r_e = 0.86$), with the electron inertial length $r_e$ = 2.8 km.

**Discussions**

The cartoon in Figure 4 summarizes the cross-scale energy transfer via multiple wave-particle resonant processes in the foreshock solar wind. The ions reflected from the terrestrial bow shock, with pitch angles of ~135° (see Figure 2a), provide the initial free energy to linearly excite the ULF waves through the classical cyclotron resonance, which is validated by a dispersion relation solver (see the positive wave growth rate in Figure s1b). As the ULF waves grow, the wave amplitude becomes comparable to the background magnetic field, which indicate that the classical theory of cyclotron resonance must be nonlinearly modified towards anomalous resonance[21]. Consequently, the ions with pitch angles between 40° and 90° are accelerated to form an unstable distribution (see Figure 2b), which in turn leads to the growth of whistler waves (see the negative energy transfer rate in Figure 3n). The whistler waves can further energize the electrons through anomalous resonance, to complete the cross-scale energy transfer from the unstable proton population to the electrons. In other words, the energy is transferred from MHD- to ion-scales, and eventually to electron-scales. These observations highlight the critical role played by wave-particle interactions in the cross-scale energy transfer, especially when a specific type of plasma waves (in this case, the whistler waves) can resonate simultaneously with different particle species. Moreover, it is found that plasma waves with the same frequency and wave number but different amplitudes during their evolution, can resonate with different plasma populations as time proceeds, which emphasize the importance of anomalous resonance in the cross-scale energy transfer at regions of weak magnetic field.

**Methods**
**Linear instability analysis**

The kinetic plasma dispersion relation solver "PDRK" [31] is applied to calculate the wave dispersion relation and the corresponding growth rate in the plasma rest frame. The input parameters derived from observations are as follows. The background magnetic field is 4nT. The plasma consists of three components, including the Maxwellian-distributed solar wind ions, electrons, and the shifted-Maxwellian reflected beam (see Figure 2a). Their temperatures are 10eV, 10eV, and 1000eV, respectively. Their number densities are 4, 4.03, and 0.03 cm$^{-3}$, respectively. The velocity shift for the beam population is 960 km/s. The parameters for the beam population are obtained by integrating the observed ion distributions within the energy range of 2 ~ 8 keV and the pitch angle range of 90° ~ 180°.

**Simulation setup**

The electromagnetic fields in the spacecraft rest frame in the FAC coordinates are expressed as,

$$\mathbf{E} = \mathbf{E_1} = E_{max} \exp\left(-\frac{(z - v_w t - Z_0)^2}{L^2}\right)[\sin(k_\parallel z - \omega t)\mathbf{x}, -\cos(k_\parallel z - \omega t)\mathbf{y}], \quad (4)$$

$$\mathbf{B} = \mathbf{B_0} + \mathbf{B_1} = B_0 \mathbf{z} + B_{max} \exp\left(-\frac{(z - v_w t - Z_0)^2}{L^2}\right)[\cos(k_\parallel z - \omega t)\mathbf{x}, \sin(k_\parallel z - \omega t)\mathbf{y}], \quad (5)$$

where $\omega$ represents the wave frequency in the spacecraft rest frame. The Gaussian-profiled wave packet propagates at the velocity of $v_w$ along the uniform background magnetic field $B_0$. The Gaussian profile is initially centered at $Z_0$ with a characteristic width $L$. The parallel wave number in the spacecraft rest frame $k_\parallel = v_w/\omega$ and the wavelength $\lambda = 2\pi/k_\parallel$ are consequently defined.

For the quasi-monochromatic ULF waves, we assume a constant $\omega_{ULF} = 0.14$ rad s$^{-1}$. Other parameters, determined to match the observations, include $B_0 = 3$nT, $B_{max} = 2$nT, $E_{max} = 0.9$ mV m$^{-1}$, $v_{ULF} = 450$ km s$^{-1}$, $k_{ULF} = 3.1 * 10^{-7}$ m$^{-1}$, $\lambda = 2.025 * 10^7$m, $Z_0 = 5\lambda$ and $L = 2\lambda$. The resonance condition in Figure s2a is obtained based on these parameters.

For the dispersive whistler waves, the situation is more complicated since the wave phase velocity depends on frequency. In this event, the difference in the wave phase velocity (due to the changing wave frequency) is ~20 km/s, which is much smaller than the typical phase velocity of ~400 km/s. Therefore, we neglect the variations of the wave phase velocity, and the wave packet hardly changes its profile within the time interval of interest. We also assume that the whistler wave frequency $\omega_{wh}$ decreases exponentially with time,

$$\omega_{wh}(z, t) = 0.2\pi \exp((v_{wh} t + Z_1 - z)/L_1) \quad (6)$$

Here, $Z_1$ represents the initial position where the wave frequency is $0.2\pi$, and $L_1$ represents the characteristic width of the frequency profile. With a fixed wave phase speed $v_w$, the parallel wave number $k_{wh}(z, t) = v_{wh}/\omega_{wh}(z, t)$ and the wavelength $\lambda_{wh}(z, t)$, are consequently determined as functions of position and time, in the spacecraft rest frame. The parameters for the whistler waves are $B_0 = 2$nT, $B_{max} = 1.5$nT, $E_{max} = 0.6$ mV m$^{-1}$, $v_{wh} = 400$ km s$^{-1}$, $Z_0 = -3.5 * 10^6$m, $L = 10^6$m, $Z_1 = -2.4 * 10^6$m and $L_1 = 2 * 10^6$m. To achieve the resonance conditions in Figures s2b-s2c, $\omega_{wh}$ is fixed at 2.8 rad s$^{-1}$.

In the simulation, the initial ion distributions are obtained by direct observations (the averaged FPI measurements between 0816:00 and 0817:00). The initial electron distributions are assumed to be shifted bi-Maxwellian, expressed as,

$$f = n_0 \left(\frac{m}{2\pi T_\perp}\right)\left(\frac{m}{2\pi T_\parallel}\right)^{\frac{1}{2}} \exp\left(-\frac{mV_x^2 + mV_y^2}{2T_\perp} - \frac{m(V_z - V_e)^2}{2T_\parallel}\right), \quad (7)$$

where the parameters include the parallel bulk velocity $V_e = 450 km\ s^{-1}$, parallel temperature $T_\parallel = 20$eV, perpendicular temperature $T_\perp = 10$eV and electron number density $n_0 = 1.8\ cm^{-3}$. These parameters are determined from the MMS FPI measurements.

The immobile virtual spacecraft is located at (0,0,0) in this simulation, which obtains the proton velocity distributions every 3 seconds. At any given time, 12 protons (evenly-distributed in the perpendicular plane, with 30° gyrophase interval) with the same kinetic energy and pitch angle are traced backward to obtain their initial positions in the velocity phase space. The phase space density for each bin is consequently determined by tracing its corresponding particle via the Liouville's theorem. For the electrons, the time resolution is 0.06s, and the gyrophase interval is 10°.

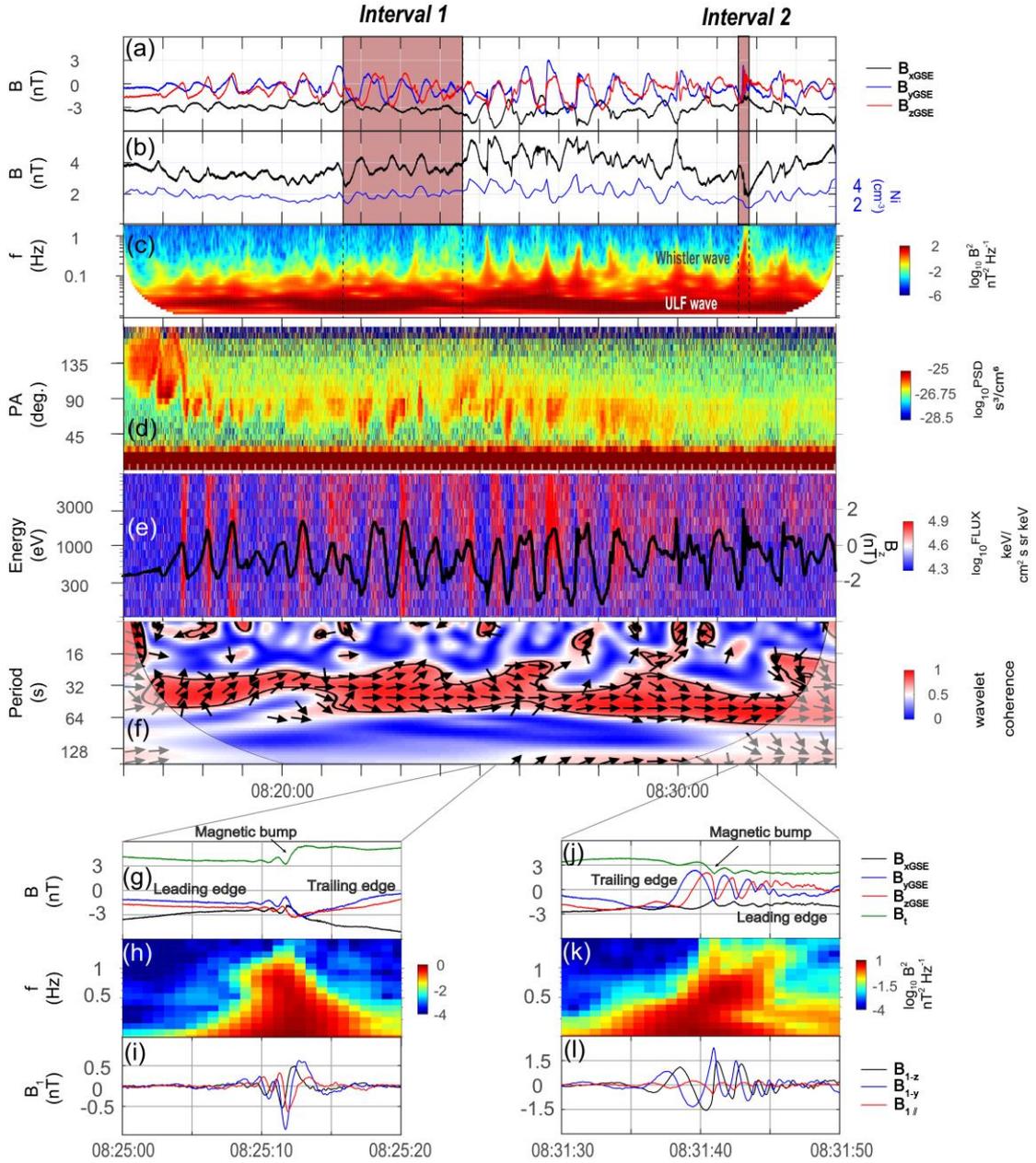

Figure 1. MMS observations of ULF waves, ion distributions and whistler waves. (a) Magnetic field in the GSE coordinates. (b) Magnetic field strength, with the ion number density superimposed. (c) Wavelet power spectral density of the perpendicular magnetic field. (d) Pitch-angle spectra of ions within the energy range between 1.5 and 4.5 keV. (e) Energy spectrum for the ions along the -z direction in the GSE coordinates, with $B_z$ component of magnetic field superimposed. (f) Wavelet coherence between the flux of the 3 keV ions moving in the -z direction and magnetic field $B_z$ component. The arrows represent their phase differences, with rightward arrows indicating the in-phase relationship. (g-i) The magnetic field, wavelet power spectra, and the filtered wave magnetic field from 0825:00 to 0825:20. Panels (j-l) are in the same format as panels g-i but from 0831:30 to 0831:50. The two shadowed intervals are selected to analyze the wave-particle interactions.

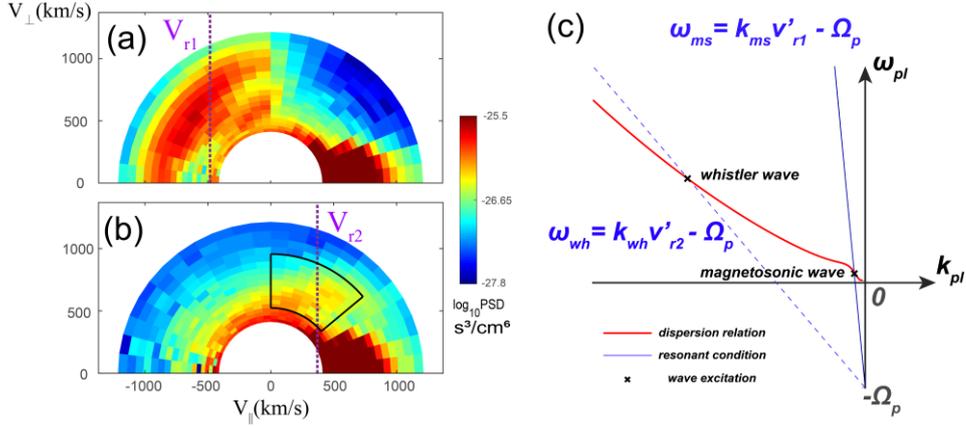

Figure 2. The observed ion distributions in the $v_\parallel - v_\perp$ plane of the spacecraft frame and the corresponding resonance conditions. (a) Ion distributions averaged from 0816:00 to 0817:00. The dashed purple line delineates the proton cyclotron resonance velocity $V_{r1}$ for small-amplitude ULF waves. (b) Ion distributions averaged from 0831:30 to 0831:50. The area surrounded by the solid lines represents the phase-bunched ions in resonance with the large-amplitude ULF waves. The dashed purple line delineates the proton cyclotron resonance velocity $V_{r2}$ for whistler waves. (c) Illustration of the wave dispersion relation (red) and wave-particle resonance conditions (blue) in the plasma rest frame. Their intersections indicate the frequency and wavenumber of the waves to be excited. Here, the $V_{r1}'$ and $V_{r2}'$ values are $V_{r1}$ and $V_{r2}$ transformed to the plasma rest frame.

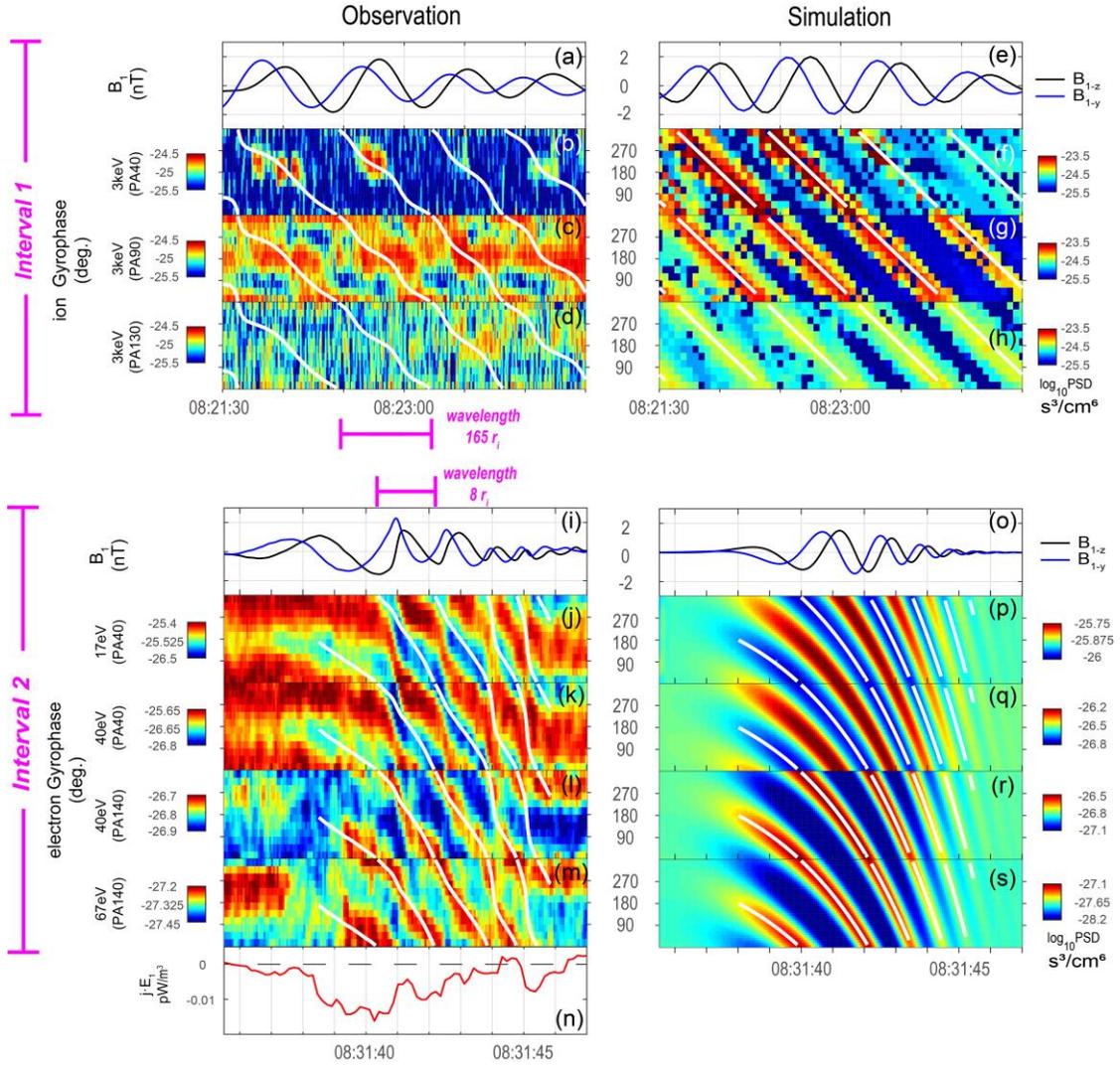

Figure 3. MMS observations and simulation results of the particle gyrophase spectra during the two shadowed intervals in Figure 1. (a) The filtered ULF wave magnetic field during the first shadowed interval. (b-d) Gyro-phase spectra for 3keV ions, at the pitch angles of 40°, 90°, and 130°, respectively. The phase of $-\mathbf{B}_{1,\mathbf{ULF}}$ is indicated by the white lines. (e-h) Simulation results in the same format as the observations in panels a-d. (i) The filtered whistler wave magnetic field during the second shadowed interval. (j-k) Gyro-phase spectra for electrons with the pitch angle of 40°, within the 17eV and 39eV energy channels. (l-m) Gyro-phase spectra for electrons with the pitch angle of 140°, within the 39eV and 67eV energy channels. The phase of $-\mathbf{B}_{1,\mathbf{wh}}$ is indicated by the white lines. (n) Energy transfer rate between the whistler waves and resonant ions. (o-s) Simulation results in the same format as in panels i-m.

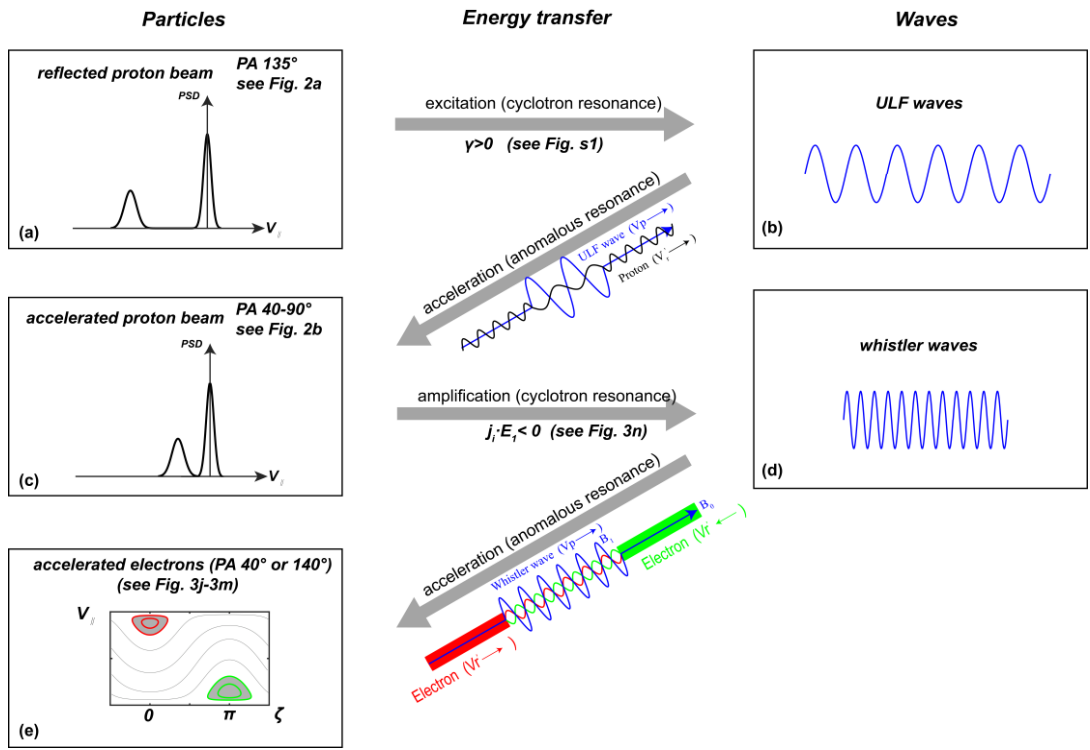

Figure 4. Illustration of the cross-scale energy transfer via wave-particle interactions.

## Data availability

All MMS data are available to the public via https://lasp.colorado.edu/mms/sdc/public/.

## Code availability

The MMS data are processed and analyzed using the IRFU-Matlab package available at https://github.com/irfu/irfu-matlab. The test-particle simulation codes are also available from Github (https://github.com/lijinghuan1997/cross-scale-shocklet).

## Acknowledgements

We are grateful to the MMS mission and team for the high-resolution measurements of the particles and fields. This work is supported by the National Natural Science Foundation of China grants 42174184.

## Author contributions

J.-H.L. analyzed the observational data, conducted the simulation and prepared the draft manuscript. X.-Z.Z. oversaw this project and took responsibility for the manuscript preparation. Z.-Y.L., S.W., Y.O., L.L., C.Y. and Q.-G.Z. contributed to the data interpretation and paper writing. G.L., C.T.R., and J.L.B. assured the MMS data quality. All authors contributed to the discussion of the results.

## Competing interests

# Supplementary Information for

"Direct observations of cross-scale energy transfer in space plasmas"

Jing-Huan Li[1], Xu-Zhi Zhou[1*], Zhi-Yang Liu[1], Shan Wang[1], Yoshiharu Omura[2],
Li Li[1], Chao Yue[1], Qiu-Gang Zong[1], Guan Le[3], Christopher T. Russell[4], James L. Burch[5]

6. School of Earth and Space Sciences, Peking University, Beijing, China
7. Research Institute for Sustainable Humanosphere, Kyoto University, Kyoto, Japan.
8. NASA Goddard Space Flight Center, Greenbelt, MD, USA.
9. Institute of Geophysics and Planetary Physics, University of California, Los Angeles, CA, USA.
10. Southwest Research Institute, San Antonio, TX, USA.

* Corresponding author: Xu-Zhi Zhou (xzzhou@pku.edu.cn)

This file includes Supplementary Figure 1-3.

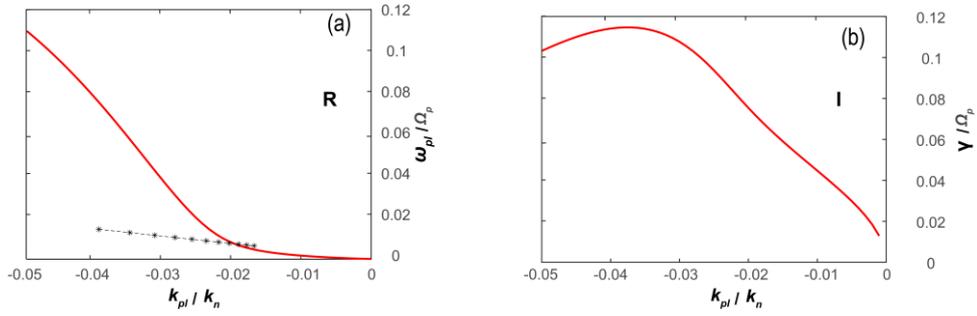

**Supplementary Figure 1**. The linear instability analysis based on the PDRK dispersion relation solver. (a) Dispersion relation. (b) Wave growth rate. The observed wave properties ($\omega_{pl,ULF}$ and $k_{pl,ULF}$ in the plasma rest frame, varied with time) are shown as the star line. The frequency is normalized by the proton cyclotron frequency $\Omega_p$, and the parallel wave number is multiplied by the ion inertial length.

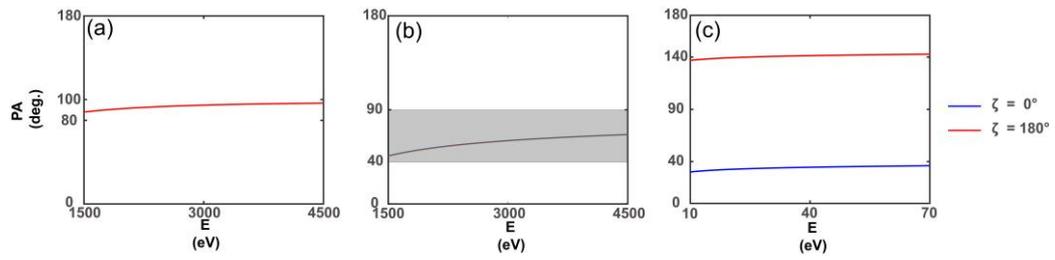

**Supplementary Figure 2.** Various resonance conditions. (a) Resonance condition ($\zeta = 180°$) for the protons interacting with the large-amplitude ULF waves. (b) Resonance condition for the protons interacting with the whistler waves ($\omega_{wh} = 2.8$ rad s$^{-1}$). (c) Resonance conditions ($\zeta = 0°/180°$, blue/red) for the electrons interacting with the whistler waves ($\omega_{wh} = 2.8$ rad s$^{-1}$).

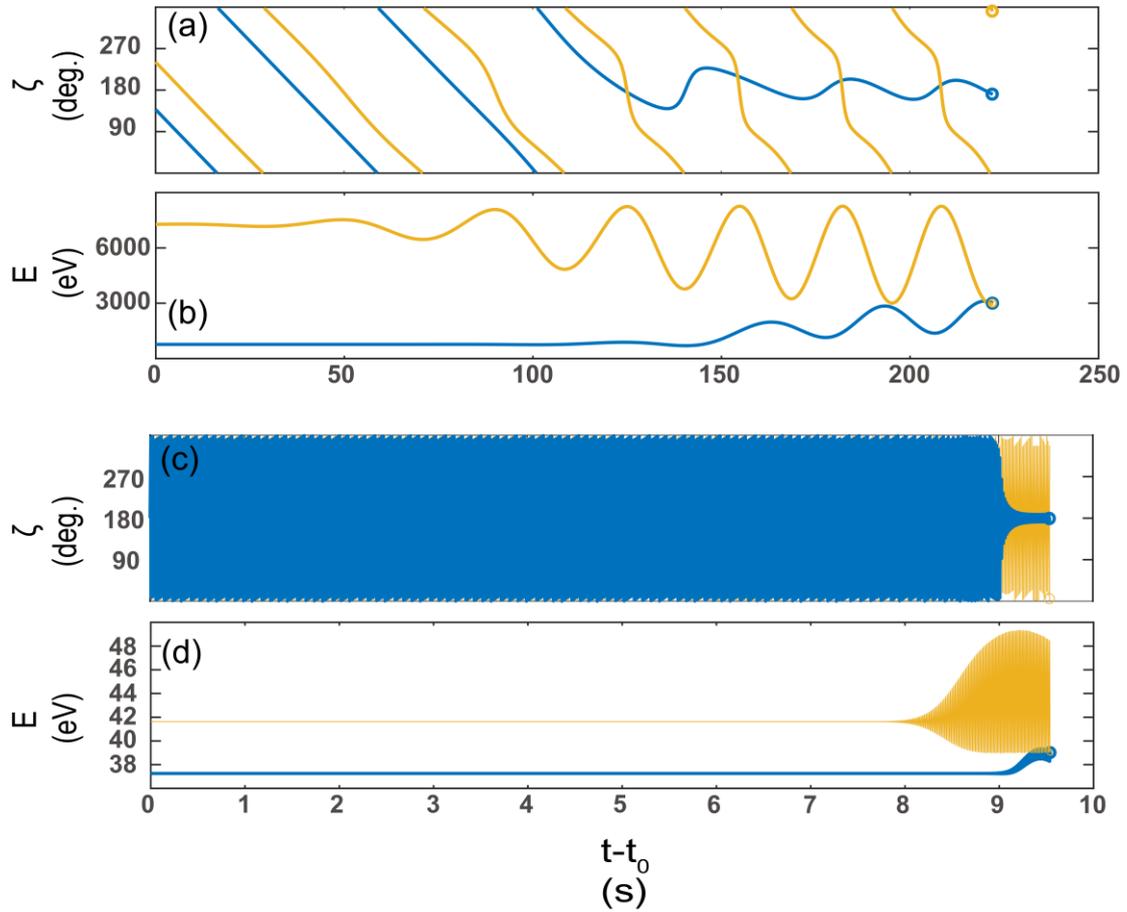

**Supplementary Figure 3**. Typical particle trajectories. (a-b) Temporal variations of $\zeta$ and kinetic energy for sample protons. (c-d) Temporal variations of $\zeta$ and kinetic energy for sample electrons. The blue proton and electron belong to the phase-bunched population with detected $\zeta$ (when reaching the virtual spacecraft) around 180°, while the yellow particles are in the opposite direction. The detected energy and pitch angle for the protons are 3keV and 80°. The detected energy and pitch angle for the electrons are 39eV and 140°.